\documentclass[aps,preprintnumbers,amsmath,amssymb,prd,superscriptaddress,longbibliography,twocolumn]{revtex4-2}
\usepackage{dcolumn}
\usepackage{color}
\usepackage[caption=false]{subfig}
\usepackage{feynmp}
\usepackage{feynmp-auto}
\usepackage{float}
\usepackage{bbm}
\usepackage{empheq}
\usepackage{bm}
\graphicspath{ {/Users/naichuked/Library/CloudStorage/Dropbox/Z_4 anisotropy RG/Image/} }

\def\comment#1{}

\newcommand{\nc}{\newcommand}
\nc{\beq}{\begin{eqnarray}}
	\nc{\eeq}{\end{eqnarray}}
\nc{\scs}{\scriptstyle}
\nc{\setval}{\fmfset{wiggly_len}{3mm} \fmfset{arrow_len}{1.5mm}
	\fmfset{arrow_ang}{13} \fmfset{dash_len}{1.5mm}\fmfpen{0.125mm}
	\fmfset{dot_size}{2thick}}

\usepackage{bm,latexsym,mathrsfs,enumerate,color}
\usepackage[mathcal]{euscript}
\usepackage[breaklinks=true,unicode=true,urlcolor = blue,colorlinks = true,citecolor = blue,linkcolor = blue]{hyperref}
\usepackage{graphicx}
\usepackage{todonotes}
\usepackage{wrapfig}

\def\slashchar#1{\setbox0=\hbox{$#1$}           
	\dimen0=\wd0                                 
	\setbox1=\hbox{/} \dimen1=\wd1               
	\ifdim\dimen0>\dimen1                        
	\rlap{\hbox to \dimen0{\hfil/\hfil}}      
	#1                                        
	\else                                        
	\rlap{\hbox to \dimen1{\hfil$#1$\hfil}}   
	/                                         
	\fi}                                         %

\DeclareMathAlphabet\mathbfcal{OMS}{cmsy}{b}{n}

\usepackage{physics}
\usepackage{xcolor}

\usepackage{multirow}
\usepackage{hhline}

\begin{document}

\title{Real critical exponents from the $\varepsilon$-expansion in an interacting $U(1)$ model with non-Hermitian $Z_4$ anisotropy}

\author{Eduard Naichuk}
\affiliation{Institute for Theoretical Solid State Physics, IFW Dresden, Helmholtzstr. 20, 01069 Dresden, Germany}
\affiliation{Bogolyubov Institute for Theoretical Physics, 03143 Kyiv, Ukraine}

\author{Jeroen van den Brink}
\affiliation{Institute for Theoretical Solid State Physics, IFW Dresden, Helmholtzstr. 20, 01069 Dresden, Germany}
\affiliation{Institute for Theoretical Physics and W\"urzburg-Dresden Cluster of Excellence ct.qmat, TU Dresden, 01069 Dresden, Germany}

\author{Flavio S. Nogueira}
\affiliation{Institute for Theoretical Solid State Physics, IFW Dresden, Helmholtzstr. 20, 01069 Dresden, Germany}

\date{\today}

\begin{abstract}
In quantum optics and condensed matter physics non-Hermitian phenomena are often studied under the assumption of an open physical system. However, there are many examples of intrinsically non-Hermitian, though often $\mathcal{PT}$-symmetric systems, which do not necessarily need to characterized as open, where it one usually speak of gain and loss relative an underlying environment. A particularly intriguing example with experimental consequences in the literature is QCD at finite density. Motivated by the existence of such inherently non-Hermitian systems, here we study the critical behavior of a $U(1)$-invariant Lagrangian perturbed by a complex, $\mathcal{PT}$ symmetric $Z_{4}$ anisotropy.  We find real critical exponents both in the region of unbroken and broken $\mathcal{PT}$ symmetry. In the former the coupling constants for fixed points or lines are real, whereas in the latter they become complex. Importantly, the most stable fixed point corresponds to the flow at large distances towards an effectively Hermitian $U(1)$ symmetric system. This constitutes an example where both the $U(1)$ and the Hermitian character are emergent features of the theory. This tells us about the importance and physical meaning of some non-Hermitian systems beyond interpretations involving gain and loss.  
\end{abstract}

\maketitle

\section{Introduction and motivation}

The textbook paradigm for quantum mechanical systems posits that their physical properties are described by Hermitian Hamiltonians. This immediately implies a unitary time evolution of the system and that energy eigenvalues are always real numbers. On the other hand, it is well known that real energy eigenvalues can also arise in non-Hermitian Hamiltonians \cite{Ashida}, provided one imposes the condition that the parity-time-reversal $\mathcal{PT}$ symmetry is preserved, $H=H^{\mathcal{PT}}=(\mathcal{PT})^{-1}H(\mathcal{PT})$ \cite{Bender_1998, Bender_2005}, where here $\mathcal{P}$ and $\mathcal{T}$ are the parity and time-reversal operators, respectively. Since the operator $\mathcal{PT}$ is not linear, eigenstates of $H$ may or may not be eigenstates of $\mathcal{PT}$. If every eigenfunction of a $\mathcal{PT}$-symmetric Hamiltonian is also an eigenfunction of the $\mathcal{PT}$ operator, then the $\mathcal{PT}$ symmetry of H is unbroken and all eigenvalues of $H$ are real. But if some of the eigenfunctions of a $\mathcal{PT}$-symmetric Hamiltonian are not simultaneously eigenfunctions of the $\mathcal{PT}$ operator, the $\mathcal{PT}$ symmetry is broken and some of eigenvalues of $H$ can be complex \cite{Bender_1998, Bender_2005}. In this case, we can identify the region of broken $\mathcal{PT}$ symmetry with the region of non-unitarity behavior of some model.

In quantum field theory, one of the simplest examples of an emergent non-Hermitian $\mathcal{PT}$-symmetric system is provided by the Lee model \cite{Bender_Brandt}, which was originally proposed by T. D. Lee in 1954 \cite{Lee} as a simplified example of renormalized quantum field theory (see also Chap. 12 in the textbook \cite{Barton}). The Lee model is an exactly solvable, featuring a renormalized coupling constant that upon crossing a certain critical value causes its bare counterpart to become purely imaginary, leading in this way to a non-Hermitian, but $\mathcal{PT}$-symmetric theory \cite{Bender_Brandt}. 

Another instance of an emergent complex coupling constant occurs in some theories exhibiting a so called fixed point collision \cite{Nogueira, Herbut, Nahum, Ihrig, Ma_He, Ma_Wang}. We can give an interpretation of this behavior in terms of a non-Hermitian regime in one of these theories, namely, the Abelian Higgs model in dimensions $2<d<4$ with a global $U(n)$ symmetry. It has been known for a quite long time that for $n$ below a certain critical value $n_c$ the Higgs coupling flows to complex conjugated fixed point values \cite{Halperin} (for a more recent four-loop analysis, see Ref.~\cite{Ihrig} and references therein), which was interpreted in the past as evidence for a fluctuation-induced weakly first-order phase transition \cite{Halperin}. However, we could also imagine two critical surfaces in RG space corresponding to actions featuring complex conjugated Higgs couplings. Further, we can see how this translates into a picture involving ``gain and loss" in an RG sense. Specifically, let us recall (in a somewhat schematic way) some results supporting this point of view from Ref.~\cite{Nogueira}. Assuming $n<n_c$, the complex couplings $g_\pm=g_c\pm i\varepsilon\sqrt{|\Delta|}/[2n(n+4)]$, where $\varepsilon=4-d$ and $g_c\sim\mathcal{O}(\varepsilon)$ (the precise expressions of $g_c>0$ and $|\Delta|$ in terms of $n$ are not important for our discussion), and $|\Delta|$ vanishes for $n=n_c$, thus leading to a fixed point collision. This collision leads to renormalized mass scale as a function of the dimensionless Higss coupling, $m(g)\sim \Lambda e^{-{\rm const}/(g-g_c)}$, where $\Lambda$ is a UV cutoff \cite{Nogueira}. Thus, $m\to 0$ when $g$ approaches $g_c$ from above, but it diverges when approaching $g_c$ from below. The diverging mass as $g\to g_c-$ can be interpreted as a gain while its vanishing as $g\to g_c+$ can be viewed as a loss. 

In the example of the Abelian Higgs model in $2<d<4$ the emergenece of complex conjugated couplings does not lead to real critical exponents. Instead the system faces a runaway RG flow leading to a weakly first-order phase transition \cite{Halperin}. Real exponents arise only when the fixed points collide as $n\to n_c$, leading to a phenomenon known as pseudo-criticality or, more precisely, a so called ``walking" of the RG flow \cite{Gorbenko_1}. Such a behavior occur as some generic parameter, $\alpha$, immune to the RG flow, is varied. In the example above the protected parameter is $n$. While the Abelian Higgs model leads to complex conjugated fixed points for $n<n_c$, its Lagrangian by itself is Hermitian. In this sense an interpretation analogous to the Lee model fails, since in that case a critical renormalization scale is what causes the bare coupling to become complex \cite{Bender_Brandt}.   

The typical situation considered in many works dealing with non-Hermitian phenomena, especially in quantum optics and condensed matter physics (for a review, see Ref.~\cite{Ashida}), assumes that the system in question is an open one, and we often speak of gain and loss relative to the contact with the environment \cite{bender2019pt}. However, there are examples of intrinsically non-Hermitian quantum systems where such an interpretation does not necessarily need to be invoked. An important example having an experimental impact concerns effective field theories of QCD at finite density \cite{Ogilvie,Nussinov-PRL-2025}. In this case a chemical potential is included in the QCD (or an effective version of it) Lagrangian, rending it complex, in addition of breaking Lorentz invariance. 

So far the examples we chose to mention in this introduction, refers either to an emergent non-Hermitian behavior, or those where the underlying Lagrangian is non-Hermitian. We could therefore also inquire whether a Hermitian-like behavior could emerge by quantum or thermal fluctuations following from a non-Hermitian Lagrangian. An example having this property was considered by Fisher \cite{ME_Fisher_PhysRevLett.40.1610}, long before non-Hermitian $\mathcal{PT}$-symmetric systems were discussed. It was found that a $\phi^3$ field theory with an imaginary coupling exhibits real critical exponents and correctly describes the critical Yang-Lee edge singularities. The physics here is of critical flutuations of an Ising model in an external magnetic field. This leads to physically meaningful results, despite an inherently non-unitary theory. In fact, in real time the $\phi^3$ theory with an imaginary coupling constant is $\mathcal{PT}$-symmetric, with the parity transformation being relazed as $\phi\to -\phi$. Interestingly, this problem is also related to the Reggeon field theory \cite{Cardy_2021}, whose critical behavior is quite well studied \cite{Cardy_1980}.
Motivated by this type of question, here we consider the effect of a non-Hermitian $Z_4$ (4-state clock) anisotropy in an otherwise $U(1)$ invariant interacting theory. In the case where the amplitude fluctuations of the complex order parameter field are frozen, this precisely corresponds to  
a $XY$ model with a non-Hermitian 4-state clock anisotropy, which we have considered in previous paper \cite{Naichuk_PhysRevB.110.224505}. The phase structure of its Hermitian counterpart is well known in $d=2$ \cite{Jose} and has been considered more recently in $d=3$ \cite{Hove-Sudbo_PhysRevE.68.046107,Sandvik_PhysRevLett.124.080602}. Our recent study of the non-Hermitian model showed that a fixed point collision occurs as the spacetime dimension $d\to 2$ from above, leading to a scaling behavior quite distinct from the non-Hermitian sine-Gordon model \cite{NH-sine-Gordon}, where the scalar field is non-periodic, in contrast to the non-Hermitian clock model. We have interpreted the scaling regime of the model as one exhibiting walking behavior \cite{Gorbenko_1} with respect to the dimensionality $d$. However, the RG analysis carried in Ref.~\cite{Naichuk_PhysRevB.110.224505} is not well suited to study the case $d=3$, although we have considered a $d$-dimensional RG analysis. The lack of amplitude fluctuations requires a more refined study of the phase fluctuations, which are well known to be notoriously difficult in dimensions higher than two for Lagrangians featuring a gradient term $\sim(\partial_\mu\theta)^2$, where $\theta$ is the phase of the order parameter. The study we perform in the present work resolves this difficulty by accounting to amplitude fluctuations as well, which are encoded in the complex order parameter field $\psi$ of the theory. This immediately permits the applicability of standard $\varepsilon$-expansion techniques around the upper critical dimension.     

The non-Hermitian $Z_4$ interaction considered in this paper reads $V_{Z_{4}}(\psi)=(1/2)\left[(v+w)\psi^4+(v-w){\psi^*}^4\right]$, and is added as a perturbation to the $U(1)$ invariant $|\psi|^4$ Lagrangian. A precise definition of the model and a discussion of its non-Hermitian character is given in Section \ref{Sec:Model}. Interestingly, in this case we identify an RG invariant $k$ which is not obviously occurring in the initial action of the theory, like the case of $n$ (the number of complex field components mentioned earlier) or $d$ (the number of spacetime dimensions). The parameter in question is related to a renormalized counterpart of the ratio $v/w$ between the bare couplings $v$ and $w$. Whether the $\mathcal{PT}$ symmetry is broken or not is determined by the value of the ratio $v/w$. As it will be discussed in Section \ref{Sec:Model}, the $\mathcal{PT}$ symmetry is unbroken for $v>w$, being broken otherwise. In Section \ref{Sec:RG} we perform an RG analysis of the model and study its critical behavior using the $\varepsilon$-expansion up to $\mathcal{O}(\varepsilon^2)$. The RG invariant $k$ leads to the appearance of not only fixed points, but also fixed lines (i.e., lines of fixed points). Tuning $k$ allows to go from the regime where the $\mathcal{PT}$ symmetry is preserved to one where it is broken. In the latter case fixed points and lines become complex-valued. Remarkably, we find real, $k$-independent critical exponents in both regimes. The universality class involves critical regimes governed by Heisenberg, Ising, and cubic behaviors like in systems with cubic anisotropy \cite{Zinn_Justin}, with the most stable one being the $U(1)$ Heisenberg critical regime. Since the latter does not involve the non-Hermitian $Z_4$-symmetric interaction, we have here an example of both emergent $U(1)$ symmetry and Hermitian system.

\section{Model}
\label{Sec:Model}

In this work we study the critical behavior of the following $d$-dimensional complex scalar $U(1)$-invariant model,
\begin{eqnarray}
	\label{Eq:L}
   	\mathcal{L}&=&|\partial_\mu\psi|^2+m^2|\psi|^2+\frac{u}{2}|\psi|^4+V_{Z_{4}}(\psi),
\end{eqnarray}
where a $Z_{4}$ anisotropy potential is included, 
\begin{eqnarray}
	\label{Eq:V_Z_4}
   	V_{Z_{4}}(\psi)&=&\frac{1}{2}\left[(v+w)\psi^4+(v-w){\psi^*}^4\right],
\end{eqnarray}
To ensure the stability of the real part of the potential in the Lagrangian \eqref{Eq:L}, we require $u>2|v|$ and $v,w\in\mathbb{R}$. The non-Hermitian character of the theory is encoded in the terms having a nonzero value of the coupling constant $w$. Following other studies of non-Hermitian field theories \cite{Ogilvie,Nussinov-PRL-2025,ME_Fisher_PhysRevLett.40.1610}, we also consider an Euclidean formulation for our model.

Since $\mathcal{PT}$ symmetry maps both $\psi$ and $\psi^*$ into $-\psi$ and $-\psi^*$ and $\mathcal{T}i\mathcal{T}^{-1}=-i$, the corresponding Hamiltonian is also $\mathcal{PT}$-symmetric. The Hermitian model $(w=0)$ applies to different physical contexts. For instance, it occurs as an effective theory associated to the breaking of the $SU(N)$ symmetry in a non-Abelian Higgs model \cite{Hooft}. More recently, it has been studied for $d=3$ as an effective field theory for the transition to valence-bond solid phases \cite{Pujari}. The Lagrangian is closely related to the 4-state clock model. This connection is easily established by rewriting the complex scalar field in terms of polar coordinates, $\psi=\rho e^{i\theta}/\sqrt{2}$, in which case the Lagrangian becomes,
\begin{eqnarray}
	\label{Eq:L-1}
	\mathcal{L}&=&\frac{1}{2}(\partial_\mu\rho)^2+\frac{\rho^2}{2}(\partial_\mu\theta)^2+\frac{m^2}{2}\rho^2+\frac{u}{8}\rho^4\nonumber\\
	&+&\frac{\rho^4}{4}\left[v\cos(4\theta)+iw\sin(4\theta)\right]. 
\end{eqnarray} 
With this parametrization the $\mathcal{PT}$ symmetry acts as, $\mathcal{P}\theta\mathcal{P}^{-1}=-\theta+\pi$ and $\mathcal{T}i\mathcal{T}^{-1}=-i$. The amplitude $\rho$ is both $\mathcal{P}$- and $\mathcal{T}$-invariant. Reparametrizing the phase as $\theta\to\theta+(i/4){\rm arctanh}(w/v)$ \cite{Bender} leads to the following equivalent Lagrangian, 
\begin{eqnarray}
	\label{Eq:L-2}
	\mathcal{L}&=&\frac{1}{2}(\partial_\mu\rho)^2+\frac{\rho^2}{2}(\partial_\mu\theta)^2+\frac{m^2}{2}\rho^2+\frac{u}{8}\rho^4\nonumber\\
	&+&\frac{\rho^4}{4}\sqrt{v^2-w^2}\cos 4\theta.
\end{eqnarray}

Concerning the above form of the Lagrangian, some remarks are in order. Recall that we call a model $\mathcal{PT}$-symmetric if $[\mathcal{H},\mathcal{PT}]=0$, where $\mathcal{H}$ is the density of the Hamiltonian corresponding to the density of the Lagrangian $\mathcal{L}$. But since the operator $\mathcal{PT}$ is not linear, eigenstates of $\mathcal{H}$ may or may not be eigenstates of $\mathcal{PT}$. If every eigenfunction of a $\mathcal{PT}$-symmetric Hamiltonian is also an eigenfunction of the $\mathcal{PT}$ operator, then by definition the $\mathcal{PT}$ symmetry is \textit{unbroken}. Moreover, it can be verified that in this case all eigenvalues of $\mathcal{H}$ are real. However, if some of the eigenfunctions of a $\mathcal{PT}$-symmetric Hamiltonian are not simultaneously eigenfunctions of the $\mathcal{PT}$ operator, then by definition the $\mathcal{PT}$ symmetry is \textit{broken} and some of eigenvalues of $\mathcal{H}$ can be complex. In the parameterization of the Lagrangian \eqref{Eq:L-2} we see that spectrum is always real for $v^2>w^2$ and can be complex for $v^2<w^2$. As a result, we conclude that for $v^2>w^2$ we are dealing with a region of unbroken $\mathcal{PT}$ symmetry, and for $v^2<w^2$ we have a region of broken $\mathcal{PT}$ symmetry, separated by the exceptional point $v^2=w^2$. It is worth mentioning that a similar argument was applied in the context of a non-Hermitian sine-Gordon theory in Ref.~\cite{NH-sine-Gordon}.

\section{Renormalization group analysis}
\label{Sec:RG}

In order to perform the RG analysis, we decompose the complex scalar field into its real and imaginary components,  $\psi=(\phi_1+i\phi_2)/\sqrt{2}$ and rewrite the potential \eqref{Eq:V_Z_4} added to $u|\psi|^4/2$ as,  
\begin{eqnarray}
    &\frac{u}{2}|\psi|^4+V_{Z_{4}}(\phi_1,\phi_2)=\frac{1}{8}\left(u+2v\right)(\phi^4_1+\phi^4_2)
    \nonumber\\
    &+\frac{1}{4}(u-6v)\phi^2_1\phi^2_2
	+iw(\phi^3_1\phi_2-\phi^3_2\phi_1).
\end{eqnarray}
The $\mathcal{P}$ and $\mathcal{T}$ operators act on the fields $\phi_1$ and $\phi_2$ as,
\begin{subequations}
    \begin{empheq}[]{align}
        &\mathcal{P}\phi_1\mathcal{P}=\phi_1, \quad \mathcal{P}\phi_2\mathcal{P}=-\phi_2,\\ 
        &\mathcal{T}\phi_1\mathcal{T}=-\phi_1, \quad \mathcal{T}\phi_2\mathcal{T}=-\phi_2.
    \end{empheq}
\end{subequations}
The above equations imply that $\phi_1$ is $\mathcal{PT}$-odd, while $\phi_2$ is $\mathcal{PT}$-even. Thus, a $\mathcal{PT}$ symmetry breaking pattern where the expectation value $\langle\phi_1\rangle\neq 0$ leads immediately to the generation of $\phi_2^3$ term with a purely imaginary coefficitient, in a scenario that closely parallels the one of a $\phi^3$ theory with an imaginary coupling constant \cite{ME_Fisher_PhysRevLett.40.1610}, which we have mentioned in the Introduction as an example of a $\mathcal{PT}$-symmetry non-Hermitian theory with an interesting critical behavior and connection to Yang-Lee edge singularities \cite{ME_Fisher_PhysRevLett.40.1610,Cardy_2021}. As we will now see, the critical behavior of the Lagrangian \eqref{Eq:L} differs significantly from the one for the $\phi^3$ theory. In particular, our anomalous dimension turns out to be positive rather the negative; see also comments on unitarity Section IV.

Then quartic terms can then be represented in the form,
\begin{eqnarray}
	\frac{u}{2}|\psi|^4+V_{Z_{4}}(\phi_1,\phi_2)&=&\frac{1}{4!}\sum_{i,j,k,l=1}^{2}g_{ijkl}\phi_{i}\phi_{j}\phi_{k}\phi_{l},
\end{eqnarray}
where
\begin{eqnarray}
	g_{ijkl}=g_1S_{ijkl}+g_2F_{ijkl}+ig_3W_{ijkl},
\end{eqnarray}
with the tensors $S$, $F$ and $W$ defined by,
\begin{subequations}
\label{Eq:SFW}
    \begin{empheq}[]{align}
        &S_{ijkl}=\frac{1}{3}\left(\delta_{ij}\delta_{kl}+\delta_{ik}\delta_{jl}+\delta_{il}\delta_{jk}\right),\\ 
        &F_{ijkl}=\delta_{ij}\delta_{ik}\delta_{il},\\
        &W_{ijkl}=\frac{1}{4}(\delta_{ij}\delta_{ik}\varepsilon_{il}+\delta_{ik}\delta_{il}\varepsilon_{ij}+\delta_{ij}\delta_{il}\varepsilon_{ik}\nonumber\\
       &\hspace{0.9cm}+\delta_{jk}\delta_{jl}\varepsilon_{ji}).
    \end{empheq}
\end{subequations}
The old coupling constants $(u,v,w)$ are related to the new one $g=(g_1,g_2,g_3)$ in the following way $g_1=3(u-6v)$, $g_2=4!v$, and $g_3=4!w$. As a result, the $\mathcal{PT}$ symmetry is unbroken when $g_2^2>g_3^2$ and broken when $g_2^2<g_3^2$.

 \begin{figure*}[t]
 	\subfloat[Full picture of RG flows]{\includegraphics[width=0.24\linewidth]{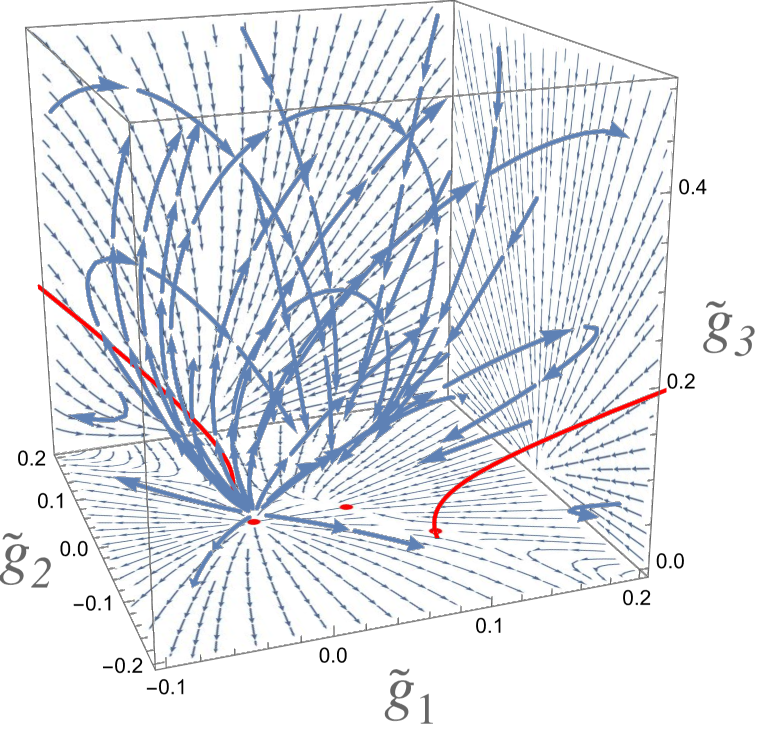}}
 	\hfill
 	\subfloat[Plane $\Tilde{g}_3=0$ ($k=0$),\newline $\mathcal{PT}$ symmetry unbroken]{\includegraphics[width=0.24\linewidth]{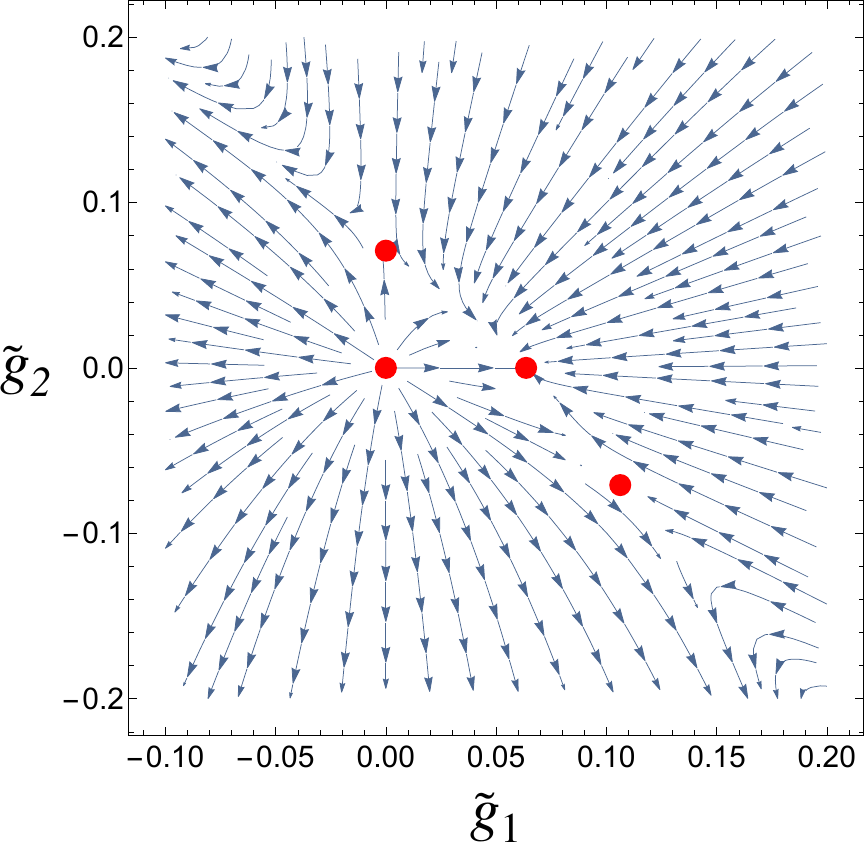}}
	\hfill
 	\subfloat[Plane $\Tilde{g}_3=0.8\Tilde{g}_2$ ($k=0.8$),\newline $\mathcal{PT}$ symmetry unbroken]{\includegraphics[width=0.24\linewidth]{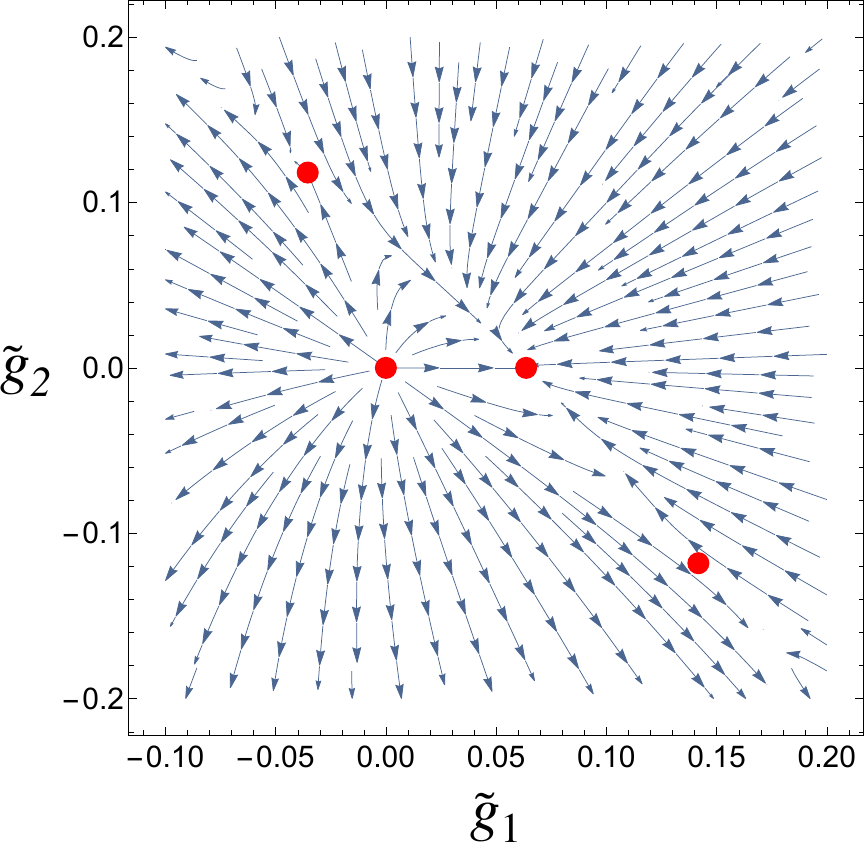}} 
	\hfill
 	\subfloat[Plane $\Tilde{g}_3=1.5\Tilde{g}_2$ ($k=1.5$), $\mathcal{PT}$ symmetry broken]{\includegraphics[width=0.24\linewidth]{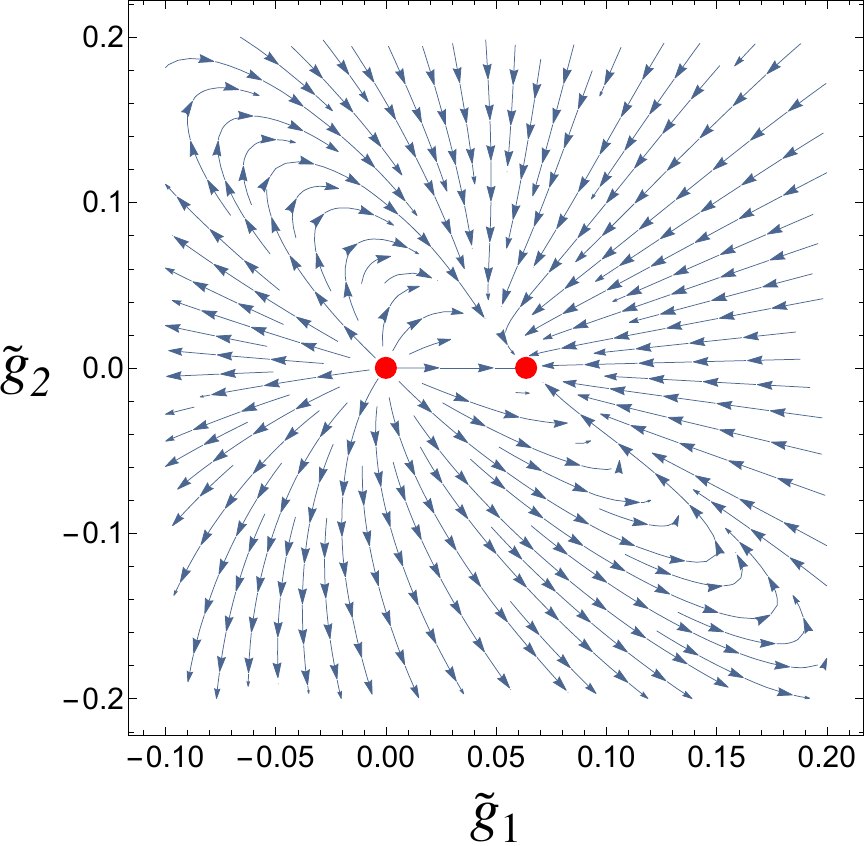}} 
 	\caption{RG flows for the non-Hermitian model associated to the Lagrangian of Eq.~\eqref{Eq:L}. The RG flows shown here are for $d=4-\varepsilon$. In all panels the red dot represents the fixed point and red line represents the fixed line. Panel (a) show full picture of RG flows. Panels (b), (c) and (d) show separately the $\Tilde{g}_3=0$, $\Tilde{g}_3=0.8\Tilde{g}_2$ and $\Tilde{g}_3=1.5\Tilde{g}_2$ planes appearing in panel (a). For panels (b) and (c), $k<1$ and the $\mathcal{PT}$ symmetry is unbroken. For panel (d), $k>1$, the fixed lines become complex (only two fixed points remain) and the $\mathcal{PT}$ symmetry is broken.}
 	\label{Fig:fig-RG_flow}
 \end{figure*}

Hence, by using dimensional regularization in the minimal subtraction scheme the RG functions are given at two-loop order by \cite{Zinn_Justin},
\begin{subequations}
\label{Eqs:RG}
    \begin{empheq}[]{align}
       &\beta_{ijkl}(g)=-\varepsilon g_{ijkl}\nonumber\\
       &\hspace{1cm}+\frac{N_d}{2}\sum_{m,n}(g_{ijmn}g_{mnkl}+\text{2 terms})\nonumber\\
       &\hspace{1cm}-\frac{N_d^2}{4}\sum_{m,n,p,q}(g_{ijmn}g_{mpqk}g_{npql}+\text{5 terms})\nonumber\\
       &\hspace{1cm}+\frac{N_d^2}{48}\sum_{m,n,p,q}(g_{ijkm}g_{mnpq}g_{npql}+\text{3 terms}), \label{Eq:beta}\\ 
       &\eta(g)\delta_{ij}=\frac{N_d^2}{24}\sum_{k,l,m}g_{iklm}g_{jklm}, \label{Eq:eta}\\
       &\eta_2(g)\delta_{ij}=-\frac{N_d}{2}\sum_{k}g_{ijkk}+\frac{5N_d^2}{24}\sum_{k,l,m}g_{iklm}g_{jklm}, \label{Eq:eta_2}
    \end{empheq}
\end{subequations}
where $N_d=2/(4\pi)^{d/2}\Gamma(d/2)$ and $\varepsilon=4-d$. The RG function $\beta_{ijkl}$ is the RG $\beta$ function for the coupling $g_{ijkl}$, $\eta(g)$ is the anomalous dimension for the two-point correlation function, and $\eta_{2}(g)$ is the anomalous dimension associated to insertions of the composite operator $(1/2)\phi_i\phi_j$. The latter determines the scaling behavior of the correlation length. The additional terms in the expressions above refer to permutations of the indices respecting the symmetry of the vertex function.  From Eq.~\eqref{Eq:beta} for $d=4-\varepsilon$, one derives the $\beta$ functions in the $\varepsilon$-expansion as, 
\begin{subequations}
\label{tilde_system_beta}
    \begin{empheq}[]{align}
        &\Tilde{\beta}_1(\Tilde{g})=-\varepsilon \Tilde{g}_1+\frac{5}{3}\Tilde{g}_1^2+\Tilde{g}_1\Tilde{g}_2-\frac{3}{16}\Tilde{g}_3^2-\frac{5}{3}\Tilde{g}_1^3\nonumber\\
        &\hspace{0.85cm}-\frac{11}{6}\Tilde{g}_1^2\Tilde{g}_2-\frac{5}{12}\Tilde{g}_1\Tilde{g}_2^2+\frac{23}{48}\Tilde{g}_1\Tilde{g}_3^2+\frac{3}{8}\Tilde{g}_2\Tilde{g}_3^2, \label{Eq:beta_1}\\ 
        &\Tilde{\beta}_2(\Tilde{g})=-\varepsilon \Tilde{g}_2+\frac{3}{2}\Tilde{g}_2^2+2\Tilde{g}_1\Tilde{g}_2-\frac{17}{12}\Tilde{g}_2^3-\frac{23}{9}\Tilde{g}_1^2\Tilde{g}_2\nonumber\\
        &\hspace{0.85cm}-\frac{23}{6}\Tilde{g}_1\Tilde{g}_2^2-\frac{1}{48}\Tilde{g}_2\Tilde{g}_3^2, \label{Eq:beta_2}\\
        &\Tilde{\beta}_3(\Tilde{g})=-\varepsilon \Tilde{g}_3+\frac{3}{2}\Tilde{g}_2\Tilde{g}_3+2\Tilde{g}_1\Tilde{g}_3-\frac{1}{48}\Tilde{g}_3^3-\frac{23}{9}\Tilde{g}_1^2\Tilde{g}_3\nonumber\\
        &\hspace{0.85cm}-\frac{17}{12}\Tilde{g}_2^2\Tilde{g}_3-\frac{23}{6}\Tilde{g}_1\Tilde{g}_2\Tilde{g}_3, \label{Eq:beta_3}
    \end{empheq}
\end{subequations}
where $\Tilde{\beta}_{\alpha}(\Tilde{g})=N_4\beta_{\alpha}(\Tilde{g})$, $\Tilde{g}_{\alpha}=N_4g_{\alpha}$ and $ \alpha=1,2,3$. 

Note that all $\beta$ functions are real despite the fact that the initial model is non-Hermitian and contains an imaginary unit in front of the constant $w=\Tilde{g}_3/(4!N_4)$. This is because all $\beta$ functions include only terms with even degrees of $\Tilde{g}_3$, while all terms with odd degrees of $\Tilde{g}_3$ are zero. In addition, this property is preserved even in the two-loop approximation, which indicates the fundamental nature of this behavior.

The above equations imply that $\ln(\Tilde{g}_2/\Tilde{g}_3)$ is an RG invariant, so we can write $\Tilde{g}_3=k\Tilde{g}_2$, where $k$ is a constant. For $k=0$ the standard results for the $\varepsilon$-expansion of a quartic interaction $O(2)$ model with cubic anisotropy are recovered. For $k\neq 0$, we obtain that $k^2<1$ corresponds to a region of unbroken $\mathcal{PT}$ symmetry $(\Tilde{g}^2_2>\Tilde{g}^2_3)$ and, respectively, $k^2>1$ to the region of broken $\mathcal{PT}$ symmetry $(\Tilde{g}^2_2<\Tilde{g}^2_3)$. Using this linear relationship between $\Tilde{g}_2$ and $\Tilde{g}_3$, we can leave only two equations, which now depend on the constant $k$,
\begin{subequations}
\label{tilde_system_beta12}
    \begin{empheq}[]{align}
        &\Tilde{\beta}_1(\Tilde{g})=-\varepsilon \Tilde{g}_1+\frac{5}{3}\Tilde{g}_1^2+\Tilde{g}_1\Tilde{g}_2-\frac{3k^2}{16}\Tilde{g}_2^2-\frac{5}{3}\Tilde{g}_1^3\nonumber\\
        &\hspace{0.85cm}-\frac{11}{6}\Tilde{g}_1^2\Tilde{g}_2+\left(\frac{23k^2}{48}-\frac{5}{12}\right)\Tilde{g}_1\Tilde{g}_2^2+\frac{3k^2}{8}\Tilde{g}_2^3, \label{Eq:beta1}\\ 
        &\Tilde{\beta}_2(\Tilde{g})=-\varepsilon \Tilde{g}_2+\frac{3}{2}\Tilde{g}^2_2+2\Tilde{g}_1\Tilde{g}_2-\left(\frac{k^2}{48}+\frac{17}{12}\right)\Tilde{g}_2^3\nonumber\\
        &\hspace{0.85cm}-\frac{23}{9}\Tilde{g}_1^2\Tilde{g}_2-\frac{23}{6}\Tilde{g}_1\Tilde{g}^2_2. \label{Eq:beta3}
    \end{empheq}
\end{subequations}

 \begin{figure*}[t!]
 	\subfloat[]{\includegraphics[width=0.33\linewidth]{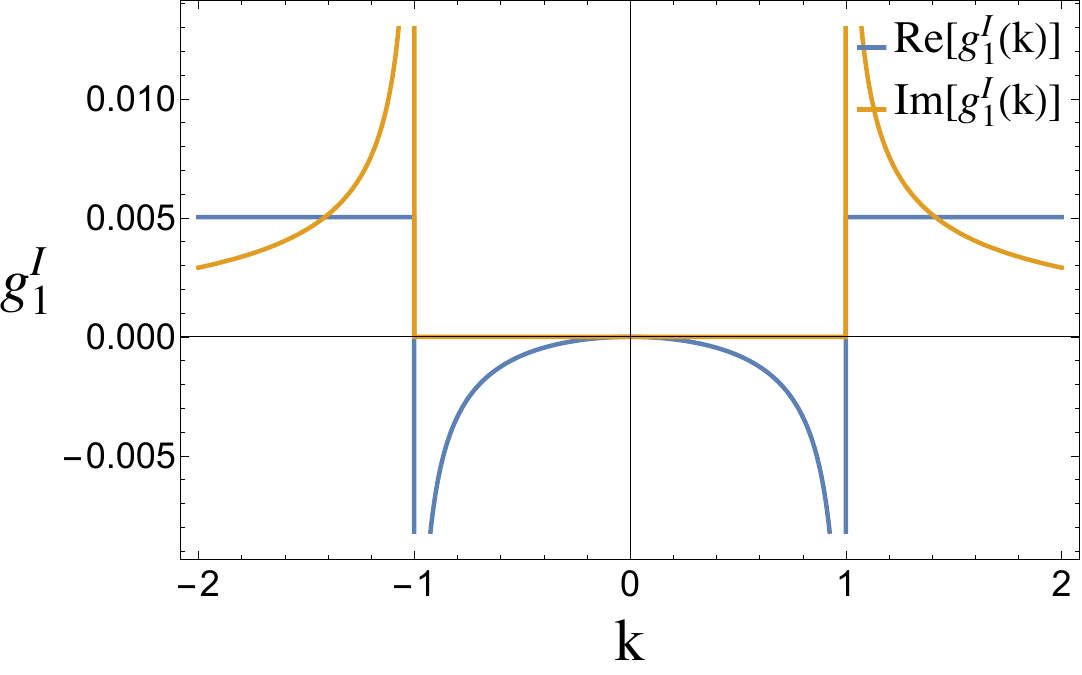}}
 	\hfill
 	\subfloat[]{\includegraphics[width=0.33\linewidth]{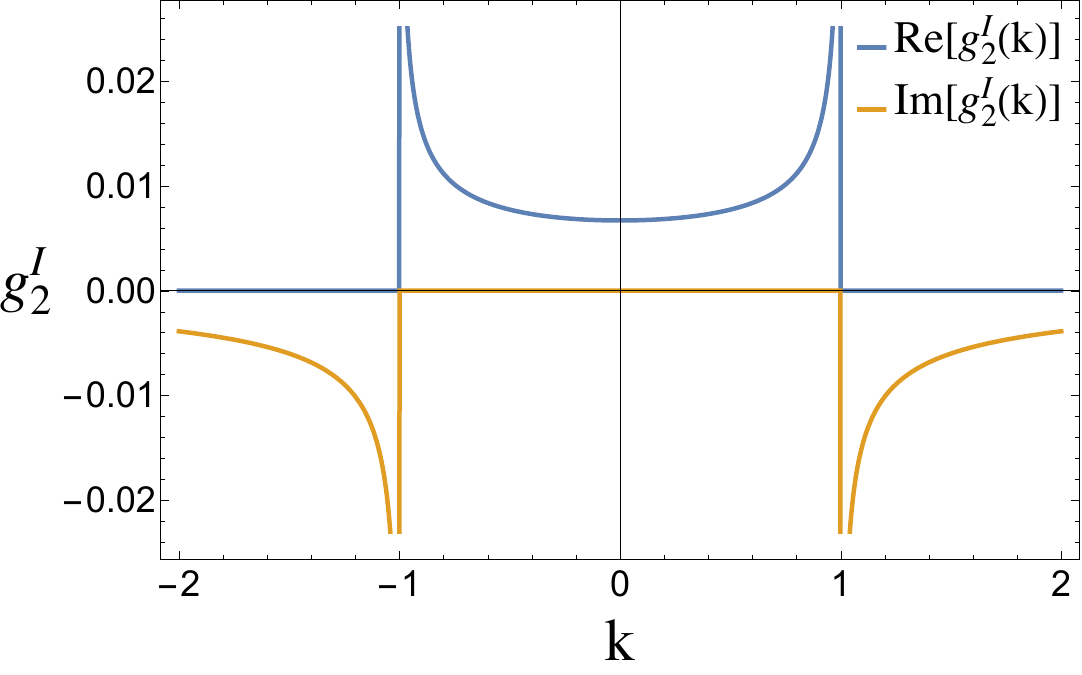}}
	\hfill
 	\subfloat[]{\includegraphics[width=0.33\linewidth]{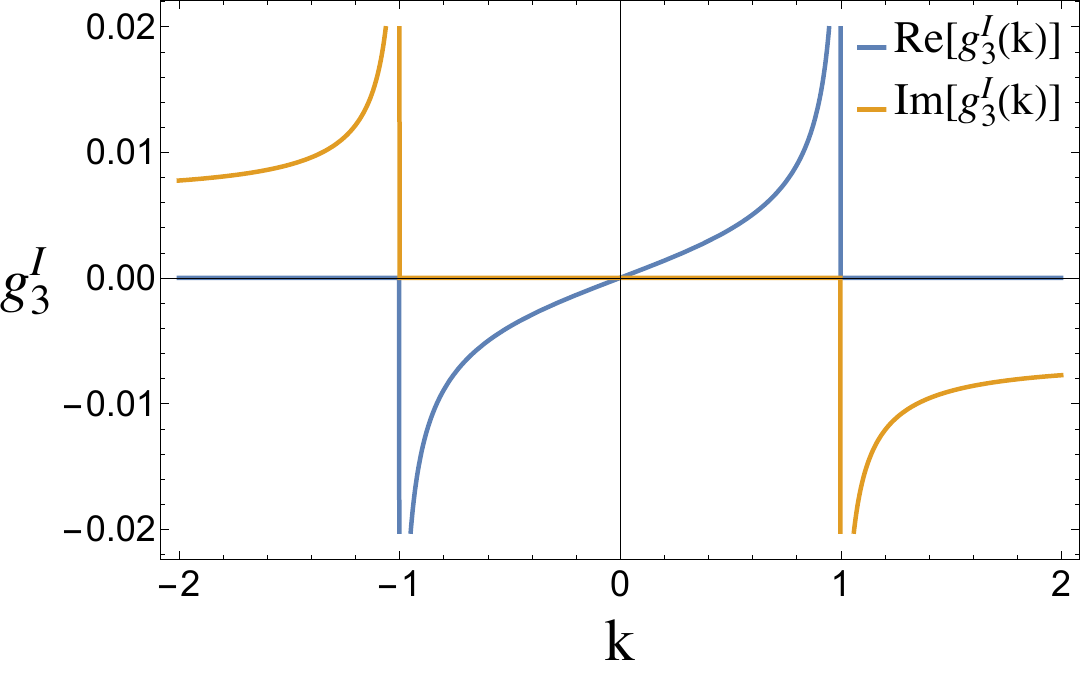}}
	\vfill
	\subfloat[]{\includegraphics[width=0.33\linewidth]{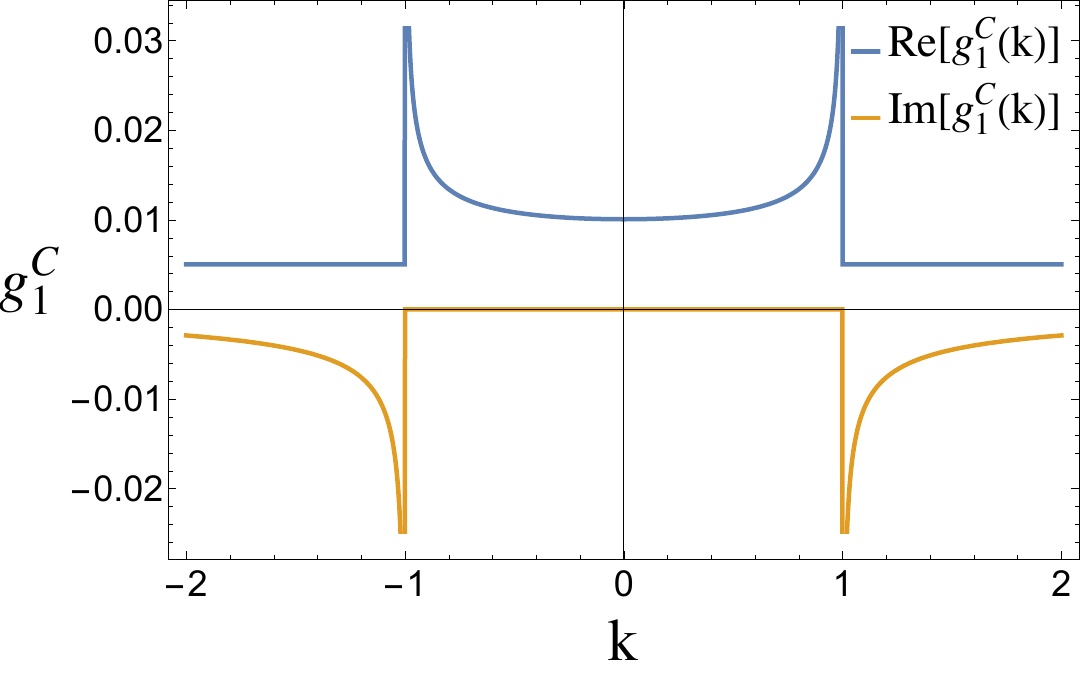}}
	\hfill
	\subfloat[]{\includegraphics[width=0.33\linewidth]{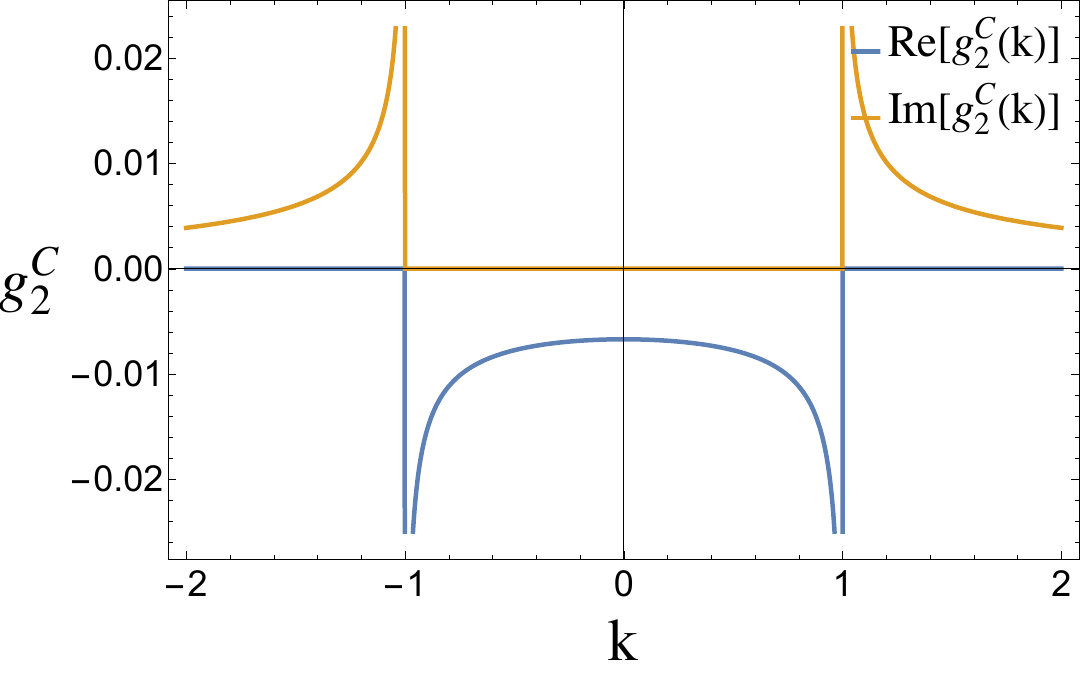}}
	\hfill
	\subfloat[]{\includegraphics[width=0.33\linewidth]{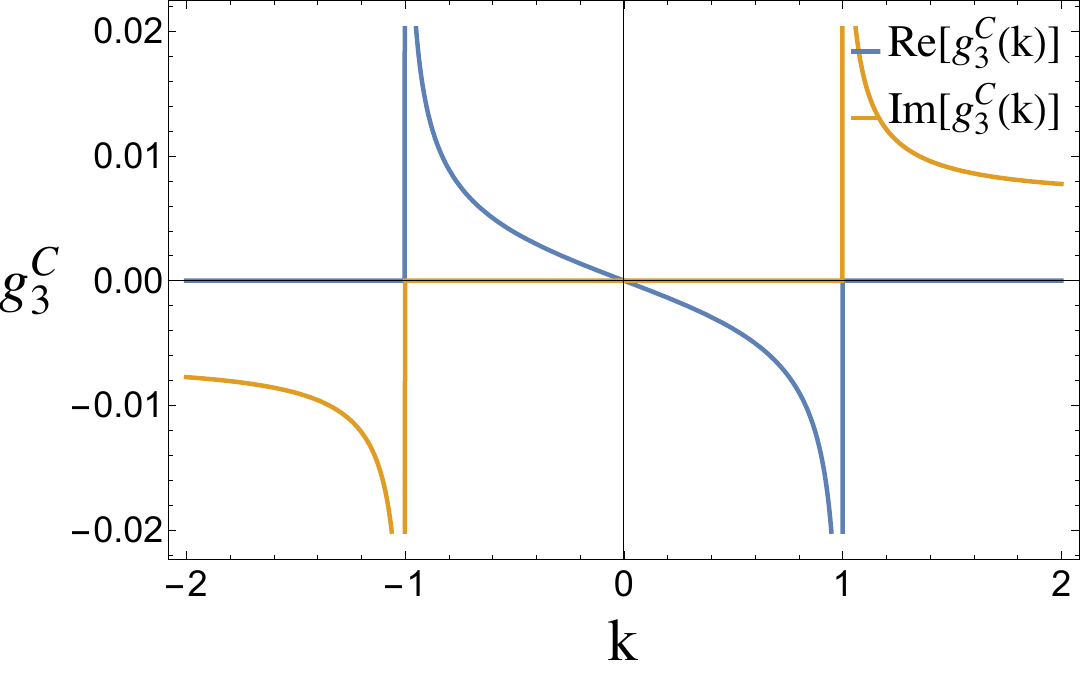}}
 	\caption{Behavior of the real and imaginary parts of the coupling constants $\Tilde{g}_1^*$, $\Tilde{g}_2^*$, and $\Tilde{g}_3^*$ for the case of Ising (a, b, c) and Cubic (d, e, f) fixed lines with respect to the parameter $k$. Region $k\in(-1,1)$ corresponds to the region of unbroken $\mathcal{PT}$ symmetry and in this region the imaginary part is always zero. Accordingly, the region $k\in(-\infty,-1)\cup(1,+\infty)$ corresponds to the region of broken $\mathcal{PT}$ symmetry and in it the imaginary part is nonzero. Points $k=\pm 1$ correspond to exceptional points, and at these points the values diverge.} 
 	\label{Fig:Re_Im}
 \end{figure*}

Clearly, scale invariance appears at the fixed points, at which all $\beta$'s vanish. We find the $\beta$ function zeros in the form of  $\varepsilon$ expansions. Details are presented in Appendix A. Besides the Gaussian fixed point $(\Tilde{g}_1^*=\Tilde{g}_2^*=\Tilde{g}_3^*=0)$ there is one nontrivial Heisenberg fixed point $(\Tilde{g}_1^*=\frac{3}{5}\varepsilon+\frac{9}{25}\varepsilon^2$, $\Tilde{g}_2^*=\Tilde{g}_3^*=0)$, generalized Ising fixed line $(\Tilde{g}_1^*=g_1^I(k)$, $\Tilde{g}_2^*=g_2^I(k)$, $\Tilde{g}_3^*=g_3^I(k))$ and generalized Cubic fixed line $(\Tilde{g}_1^*=g_1^C(k)$, $\Tilde{g}_2^*=g_2^C(k)$, $\Tilde{g}_3^*=g_3^C(k))$, shown in Fig.~\ref{Fig:fig-RG_flow} (a), which are parameterized by a constant $k$,
\begin{subequations}
\label{gI}
    \begin{empheq}[]{align}
        &g_1^I(k)=\left(\frac{\varepsilon}{2}+\frac{17\varepsilon^2}{54}\right)\left(1-\frac{1}{\sqrt{1-k^2}}\right),\\
        &g_2^I(k)=\frac{2\varepsilon}{3\sqrt{1-k^2}}+\frac{34\varepsilon^2}{81\sqrt{1-k^2}},\\
        &g_3^I(k)=\frac{2k\varepsilon}{3\sqrt{1-k^2}}+\frac{34k\varepsilon^2}{81\sqrt{1-k^2}},
    \end{empheq}
\end{subequations}
and
\begin{subequations}
\label{gC}
    \begin{empheq}[]{align}
        &g_1^C(k)=\left(\frac{\varepsilon}{2}+\frac{17\varepsilon^2}{54}\right)\left(1+\frac{1}{\sqrt{1-k^2}}\right),\\
        &g_2^C(k)=-\frac{2\varepsilon}{3\sqrt{1-k^2}}-\frac{34\varepsilon^2}{81\sqrt{1-k^2}},\\
        &g_3^C(k)=-\frac{2k\varepsilon}{3\sqrt{1-k^2}}-\frac{34k\varepsilon^2}{81\sqrt{1-k^2}}.
    \end{empheq}
\end{subequations}
As mentioned before, for $k=0$ we completely reproduce the well-known flow diagram for the quartic interaction $O(2)$ model with cubic anisotropy, shown in  Fig.~\ref{Fig:fig-RG_flow} (b) for the sake of comparison. However, for $k\neq 0$  we have a whole family of additional critical points that form  critical lines in the three-dimensional space of coupling constants, see  Figs.~\ref{Fig:fig-RG_flow} (a, c).
 
The first important result is that Ising and Cubic fixed lines extend into the complex plane for the region of broken $\mathcal{PT}$ symmetry ($k^2>1$, $\Tilde{g}^2_2<\Tilde{g}^2_3$). At the exceptional point ($k=1$, $\Tilde{g}^2_2=\Tilde{g}^2_3$), these two solutions diverge in the directions of plus and minus infinity, see Fig.~\ref{Fig:Re_Im}, and accordingly only Gaussian and Heisenberg fixed points remain Fig.~\ref{Fig:fig-RG_flow} (d). Furthermore, for values $k^2<1$, $\Tilde{g}^2_2>\Tilde{g}^2_3$, the $\mathcal{PT}$ symmetry is restored and the Ising and Cubic fixed lines pass into the real domain. The transition mechanism of real fixed points into the complex plane is well studied \cite{Wang,Nahum,Nahum_2020,Ma}, and particularly in the context of {\it walking} for complex CFTs \cite{Gorbenko_1,Gorbenko_2}. However, we are dealing here with a fundamentally novel transition mechanism. Usually, for the transition into the complex plane, we deal with fixed points collision and annihilation in the real domain. However, in our case, we first observe how the fixed points diverge from each other in the directions of plus and minus infinity. This behavior is shown in Fig.~\ref{Fig:Re_Im} for $k\in(-1,1)$. Then, after passing the exceptional point, the fixed points move into the complex plane and approach each other relative to the imaginary axis. In Fig.~\ref{Fig:Re_Im} this corresponds to the region $k\in(-\infty,-1)\cup(1,+\infty)$. Hence, we are dealing with the opposite of collision and annihilation, and therefore we would rather prefer to speak of a  "scattering" of fixed points. Since the $\beta$-functions are universal up to the order calculated here, the divergences at the fixed point values of the couplings obtained as $k\to\pm 1$ indicate the behavior of the renormalized coupling constants near exceptional points. Furthermore, $k=\pm 1$ causes the similarity transformation leading to Eq.~\eqref{Eq:L-2} to become singular. 

Similar divergent behavior near exceptional points is also observed, for example, in the generalized fidelity susceptibility of the non-Hermitian Su–Schrieffer–Heeger model \cite{Tzeng, Tzeng_2023}, in the energy-splitting susceptibility of coupled cavity systems \cite{Chen_2019} and in Petermann factors of non-Hermitian systems \cite{Wiersig_2023, Wiersig_2025}. This behavior provides a powerful tool for detecting exceptional points, which is actively discussed in the above-mentioned references.

The second important result is related to stability. To study the RG flows near the fixed point, and to determine the fixed point stability, we linearize the RG equation \eqref{tilde_system_beta} near the fixed points and fixed lines in order to calculate the stability matrix,
\begin{eqnarray}
	M_{ijkl,i'j'k'l'}=\frac{\partial\beta_{ijkl}(g^*)}{\partial g_{i'j'k'l'}}.
\end{eqnarray}
The local stability properties of fixed points are governed by the eigenvalues of this matrix. The only completely stable fixed point is the Heisenberg one. The Gaussian fixed point is completely unstable and the Ising fixed line is stable only along the $\Tilde{g}_2$ direction, while in directions $\Tilde{g}_1$ and $\Tilde{g}_3$ it is unstable. On the other hand, the Cubic fixed line is stable only in $\Tilde{g}_1$ direction, while in directions $\Tilde{g}_2$ and $\Tilde{g}_3$ it is unstable. In addition, the stability properties do not change if we move from the region of unbroken $\mathcal{PT}$ symmetry to the region where the $\mathcal{PT}$ symmetry is broken. Since in this case we are dealing with only one relevant direction for Ising and Cubic fixed lines, it can be argued that our model has greater predictive power than similar Hermitian counterparts, which are usually characterized by two relevant directions \cite{Bender_2013}.
 
Now let us move on to the analysis of critical exponents of our model. Using Eq.~\eqref{Eq:eta} we obtain,
\begin{eqnarray}
	\label{eta_3}
	\eta(\Tilde{g})&=&\frac{1}{24}\left(\frac{4}{3}\Tilde{g}_1^2+\Tilde{g}_2^2+2\Tilde{g}_1\Tilde{g}_2-\frac{1}{4}\Tilde{g}_3^2\right).
\end{eqnarray}
As a result, we can calculate the corresponding anomalous dimension,  $\eta=\eta(\Tilde{g}^*)$, for all critical points and critical lines of the model. In this case, we reproduce values well known in the literature for the $\phi^4$-model with $O(2)$ symmetric and cubic interactions: $\eta^{G}=0$, $\eta^{H}=\varepsilon^2/50$ and $\eta^{I}=\eta^{C}=\varepsilon^2/54$ \cite{Kleinert_1995}.

Another important critical exponent to consider is $\nu$, governing the divergence of the correlation length. Using Eq.~\eqref{Eq:eta_2} we obtain
\begin{eqnarray}
	\eta_{2}(\Tilde{g})&=&-\frac{1}{2}\left(\frac{4}{3}\Tilde{g}_1+\Tilde{g}_2\right)\nonumber\\
	&+&\frac{5}{24}\left(\frac{4}{3}\Tilde{g}_1^2+\Tilde{g}_2^2+2\Tilde{g}_1\Tilde{g}_2-\frac{1}{4}\Tilde{g}_3^2\right),
\end{eqnarray}
and taking into account that,
\begin{eqnarray}
	\label{nu}
	\nu(\Tilde{g})=1/(2+\eta_2(\Tilde{g})),
\end{eqnarray}
we obtain, $\nu=\nu(\Tilde{g}^*)$ for all critical points and critical lines of our model, namely, $\nu^{G}=1/2$, $\nu^H=1/2+\varepsilon/10+11\varepsilon^2/200$ and $\nu^{I}=\nu^{C}=1/2+\varepsilon/12+7\varepsilon^2/162$. These values also coincide with the values for $\phi^4$-model with $O(2)$ symmetric and cubic interactions \cite{Kleinert_1995}.

This leads us to the most surprising result following from our analysis: all critical exponents are independent of the parameter $k$, which generalizes the Ising and Cubic fixed points to fixed lines. As a result, the value of the critical exponents do not change when we move from the region of unbroken $\mathcal{PT}$ symmetry to the region of broken $\mathcal{PT}$ symmetry. This is particularly interesting, because as we move into the region of broken $\mathcal{PT}$ symmetry, the coupling constants become complex, but the critical exponents remain invariant and therefore real. This is in stark contrast with the example of the Abelian Higgs model with a global $U(n)$ symmetry for $2<d<4$ mentioned in the introduction, where complex fixed points for $n<n_c$ do not lead to a critical behavior. Instead, a weakly first-order phase transition occurs in this regime \cite{Halperin,Ihrig}.

\section{Discussion and Conclusion}
\label{Sec:Discussion}

The results obtained in this paper highlights the importance of considering  non-Hermitian systems within a context beyond the physical interpretation involving gain and loss concepts characteristic of open systems. It also shows that a Hermitian system can emerge from an initially non-Hermitian one. 

In its formulation given in Eq.~\eqref{Eq:L-2}, the non-Hermitian Lagrangian considered in this paper appears at first sight to be a generalization of the non-Hermitian, $\mathcal{PT}$-symmetric sine-Gordon Lagrangian considered in 1+1 dimensions in Refs.~\cite{Bender,NH-sine-Gordon}. However, this is not the case, since the scalar field $\theta$ is periodic, while in a sine-Gordon theory only the potential is periodic, but not the scalar field itself. For this reason, it is more accurate to regard it as a generalization of the non-Hermitian four-state clock model considered in Ref.~\cite{Naichuk_PhysRevB.110.224505}. 

The effect of amplitude fluctuations, which are absent from the analysis performed in Ref.~\cite{Naichuk_PhysRevB.110.224505}, allowed us to consider an $\varepsilon$-expansion around $d=4$, which is the upper critical dimension for the Lagrangian of Eq.~\eqref{Eq:L}. The $\varepsilon$-expansion result shows the striking insensitivity of the critical exponents to the non-Hermitian character of the theory. Indeed, the critical exponents, computed up to $\mathcal{O}(\varepsilon^2)$, are real, and do not become complex even when the $\mathcal{PT}$ symmetry is broken. 

The results presented here motivates us to inquire about other interesting examples of real critical exponents following from a non-Hermitian field theory. Most importantly, whether there is a more fundamental principle dictating how non-unitary quantum field theories analytically continued to Euclidean space may lead to real critical exponents. Regarding this point, it is worth to evoke once more the example of $\phi^3$ theory with an imaginary coupling constant \cite{ME_Fisher_PhysRevLett.40.1610} mentioned in the Introduction. In that case the negative sign of the anomalous dimension $\eta$ violates the so called infrared bound $G(p)\leq 1/p^2$ for the two-point correlation function $G(p)$ \cite{IR-bound}. The latter is intimately connected to unitarity, since it follows from the K\"all\'en-Lehmann spectral representation. A violation of the IR bound implies a violation of the latter and, therefore, of unitarity. Since at the critical point, $G(p)\sim 1/p^{2-\eta}$, a negative value of $\eta$ clearly violates the IR bound. The theory presented here, on the other hand, features a positive anomalous dimension. This raises a number of interesting questions regarding $\mathcal{PT}$-symmetric field theories, since this seems to indicate that a lack of unitarity does not violate the IR bound, making the theory more stable.

\begin{acknowledgments}
	We acknowledge financial support by the Deutsche Forschungsgemeinschaft (DFG, German Research Foundation), through SFB 1143 project A5 and the W{\"u}rzburg-Dresden Cluster of Excellence on Complexity and Topology in Quantum Matter-ct.qmat (EXC 2147, Project Id No. 390858490). 
\end{acknowledgments}

\appendix
\section{Derivation of the fixed points}
We have already emphasized that our model is characterized by an RG invariant that allows us to write $\Tilde{g}_3=k\Tilde{g}_2$, where $k$ is a constant. Therefore, to find critical points, we can move from a system of three equations to a system of two equations with two variables
\begin{subequations}
\label{Eqs_FP}
    \begin{empheq}[]{align}
        &\hspace{-0.3cm}-\varepsilon \Tilde{g}_1^*+\frac{5}{3}\Tilde{g}_1^*{}^2+\Tilde{g}_1^*\Tilde{g}_2^*-\frac{3k^2}{16}\Tilde{g}_2^*{}^2-\frac{5}{3}\Tilde{g}_1^*{}^3-\frac{11}{6}\Tilde{g}_1^*{}^2\Tilde{g}_2^*\nonumber\\
        &\hspace{0.65cm}+\left(\frac{23k^2}{48}-\frac{5}{12}\right)\Tilde{g}_1^*\Tilde{g}_2^*{}^2+\frac{3k^2}{8}\Tilde{g}_2^*{}^3=0,\\ 
        &\hspace{-0.3cm}-\varepsilon\Tilde{g}_2^*+\frac{3}{2}\Tilde{g}_2^*{}^2+2\Tilde{g}_1^*\Tilde{g}_2^*-\left(\frac{k^2}{48}+\frac{17}{12}\right)\Tilde{g}_2^*{}^3\nonumber\\
        &\hspace{0.65cm}-\frac{23}{9}\Tilde{g}_1^*{}^2\Tilde{g}_2^*-\frac{23}{6}\Tilde{g}_1^*\Tilde{g}_2^*{}^2=0.
    \end{empheq}
\end{subequations}
Since we are working in the two-loop approximation, we are looking for solutions up to the second order in $\varepsilon$
\begin{subequations}
\label{Ansatz}
    \begin{empheq}[]{align}
        &\Tilde{g}_1^*=a_1\varepsilon+b_1\varepsilon^2,\\ 
        &\Tilde{g}_2^*=a_2\varepsilon+b_2\varepsilon^2.
    \end{empheq}
\end{subequations}
Substituting ansatz \eqref{Ansatz} into Eqs.~\eqref{Eqs_FP} and collecting terms with the same exponents of the power $\varepsilon$, we have two systems of equations
\begin{subequations}
\label{Eqs_a}
    \begin{empheq}[]{align}
        &-a_1+\frac{5}{3}a_1^2+a_1a_2-\frac{3k^2}{16}a_2^2=0,\\ 
        &-a_2+2a_1a_2+\frac{3}{2}a_2^2=0,
    \end{empheq}
\end{subequations}
and
\begin{subequations}
\label{Eqs_b}
    \begin{empheq}[]{align}
        &-80a_1^3-88a_1^2a_2-20a_1a_2^2-48b_1+160a_1b_1\nonumber\\
        &\hspace{0.5cm}+48a_2b_1+48a_1b_2+23k^2a_1a_2^2+18k^2a_2^3\nonumber\\
        &\hspace{1.0cm}-18k^2a_2b_2=0,\\ 
        &-368a_1^2a_2-552a_1a_2^2-204a_2^3+288a_2b_1\nonumber\\
        &\hspace{0.5cm}+288a_1b_2+144b_2+432a_2b_2\nonumber\\
        &\hspace{1.0cm}-3k^2a_2^3=0.
    \end{empheq}
\end{subequations}
From the first system of equations \eqref{Eqs_a} we can obtain the coefficients $a_1$ and $a_2$. Next we can substitute the corresponding solutions for $a_1$ and $a_2$ into the second system of equations \eqref{Eqs_b} and obtain the coefficients $b_1$ and $b_2$. The corresponding final solutions are the well-known Gaussian and Heisenberg fixed points, and are also represented by the equations of generalize Ising \eqref{gI} and Cubic \eqref{gC} fixed lines.

\bibliographystyle{apsrev4-2}
\bibliography{Bibliography}

\end{document}